\documentclass[a4paper,prd,nofootinbib,tightenlines,twocolumn]{revtex4}

\usepackage{amsfonts,amssymb,amsthm,bbm,amsmath}
\usepackage{hyperref}

\usepackage{color,psfrag}
\usepackage[dvips]{graphicx}

\newcommand{\C}{{\mathbb C}}

\newcommand{\R}{{\mathbb R}}
\newcommand{\Z}{{\mathbb Z}}

\newcommand{\cA}{{\mathcal A}}

\newcommand{\cG}{{\mathcal G}}

\newcommand{\cM}{{\mathcal M}}

\newcommand{\cP}{{\mathcal P}}

\newcommand{\cS}{{\mathcal S}}

\newcommand{\SU}{\mathrm{SU}}

\newcommand{\SL}{\mathrm{SL}}
\newcommand{\SO}{\mathrm{SO}}
\newcommand{\U}{\mathrm{U}}

\newcommand{\be}{\begin{equation}}
\newcommand{\ee}{\end{equation}}
\newcommand{\beq}{\begin{eqnarray}}
\newcommand{\eeq}{\end{eqnarray}}
\newcommand{\bes}{\begin{eqnarray}}
\newcommand{\ees}{\end{eqnarray}}

\newcommand{\su}{{\mathfrak su}}

\newcommand{\tr}{{\mathrm{Tr}}}
\newcommand{\f}{\frac}

\newcommand{\w}{\wedge}

\def\ka{\kappa}

\newcommand{\id}{\mathbb{I}}
\def\ka{\kappa}

\def\om{\omega}

\newtheorem{theorem}{Theorem}[section]

\newtheorem{prop}[theorem]{Proposition}

\begin{document}

\title{Ashtekar-Barbero holonomy on the hyperboloid: \\
Immirzi parameter as a Cut-off for Quantum Gravity}

\author{{\bf Christoph Charles}}\email{christoph.charles@ens-lyon.fr}
\affiliation{Laboratoire de Physique, ENS Lyon, CNRS-UMR 5672, 46 all\'ee d'Italie, Lyon 69007, France}

\author{{\bf Etera R. Livine}}\email{etera.livine@ens-lyon.fr}
\affiliation{Laboratoire de Physique, ENS Lyon, CNRS-UMR 5672, 46 all\'ee d'Italie, Lyon 69007, France}

\date{\today}

\begin{abstract}

In the context of the geometrical interpretation of the spin network states of Loop Quantum Gravity, we look at the holonomies of the Ashtekar-Barbero connection on loops embedded in space-like hyperboloids. We use this simple setting to illustrate two points. First, the Ashtekar-Barbero connection is not a space-time connection, its holonomies depend on the spacetime embedding of the canonical hypersurface. This fact is usually interpreted as an inconvenience, but we use it to extract the extrinsic curvature from the holonomy and separate it from the 3d intrinsic curvature. Second, we show the limitations of this reconstruction procedure, due to a periodicity of the holonomy in the Immirzi parameter, which underlines the role of a real Immirzi parameter as a cut-off for general relativity at the quantum level in contrast with its role of a mere coupling constant at the classical level.

\end{abstract}

\maketitle

\section*{Introduction}

Loop quantum gravity (LQG) proposes a framework for a canonical quantization of general relativity reformulated as a gauge field theory. It exchanges the usual metric canonical variables, the 3-metric and its conjugate extrinsic curvature tensor, by a new canonical pair defined from the first order formulation of general relativity: the Ashtekar-Barbero connection $A$ and its conjugate triad field $E$ (for an extensive review, see \cite{thiemann}). Then one proceed to consider wave-functions of the Ashtekar-Barbero connection, $\psi(A)$, and to let geometric observables, constructed from the triad, act as differential operators. The loop quantization scheme actually relies on the choice of cylindrical functionals, which  depend of the holonomies of the connection $A$ along the edges of an arbitrary chosen graph. The sum over all graphs embedded in the canonical spatial slice is then implemented by a projective limit, which yields the Hilbert space of spin network states \cite{Ashtekar:1994mh}.

In all this construction, the Immirzi parameter $\beta$ plays a crucial role. It can be seen as a new coupling constant entering the Palatini action for general relativity in front of an almost-topological term \cite{Holst:1995pc}. But, at a deeper level, it implements a canonical transformation from the original complex self-dual Ashtekar connection, which we will call $\cA$, and the real Ashtekar-Barbero $\su(2)$-connection $A$ \cite{Rovelli:1997na}. This allowed both to work with a compact gauge group $\SU(2)$ (instead of the non-compact Lorentz group $\SL(2,\C)$) and to avoid the issue of the reality conditions\footnote{As the Ashtekar connection $\cA$ is complex and is thus not equal to its complex conjugate, it can not be simply quantized as a multiplicative operator if the scalar  product   is simply defined by the Gaussian measure. One needs to modify in a non-trivial way either the scalar product or the action of the connection operator (see e.g. \cite{Soo:2001qf}).}.
At a more effective level, it enters the loop quantum gravity dynamics in a non-trivial way and seems to be a crucial parameter in the description of quantum black holes (see the review \cite{G.:2015sda}). It also appears to control the couplings to fermionic field and possible quantum gravity induced CP violation \cite{Perez:2005pm, Freidel:2005sn, Mercuri:2006um,Mercuri:2006wb}. The main drawback of the Immirzi parameter is that the fact that Ashtekar-Barbero connection is not a space-time connection anymore and the resulting apparent loss of covariance \cite{Alexandrov:2001wt} (see also the more recent \cite{Geiller:2012dd,Achour:2013gga}). Nevertheless, this does not cause any problem in practice and one can perfectly define the kinematical Hilbert space of the theory and transition amplitudes between spin network states either by a canonical Hamiltonian \cite{Thiemann:1996aw,Bonzom:2011jv} or by a spinfoam path integral amplitude \cite{Engle:2007wy}. It can however be tempting to go back to the original complex formulation, given by the specific imaginary choice of Immirzi parameter $\beta =\pm i$, and attempt to define an analytic continuation of the real formulation of loop quantum gravity \cite{Achour:2015xga}.

Here, we would like to underline the crucial difference between the role of the Immirzi parameter at the classical level and in the quantum theory. Classically, it appears as a coupling constant in the Holst-Palatini action for the first order formulation of general relativity \cite{Holst:1995pc}. In the effective field theory paradigm, one can then investigate its renormalisation flow, together with the Newton's gravity constant and the cosmological constant, as proposed in \cite{Benedetti:2011nd,Benedetti:2011yb}. In the full quantum theory, it appears as a more essential parameter defining directly the fundamental quanta of geometry -scaling the discrete spectra of the area and volume operators- in Planck units. We would like to trace back, in the loop quantization procedure, where the Immirzi parameter acquires this deeper role.

We will use the very simple example of the holonomies of the Ashtekar-Barbero connection on a space-like 3-hyperboloid embedded in flat space-time and look at its dependence on both the hyperboloid curvature and the Immirzi parameter. As it was previously pointed out by Samuel in \cite{Samuel:2000ue}, this setting allows to illustrate that the Ashtekar-Barbero connection is not a space-time connection and how the value of a Wilson loop depends on the space-time embedding of the canonical space-like hypersurface. We will push this analysis further and illustrate the  effective compactification induced by the Immirzi parameter\footnotemark from the non-compact Lorentz group $\SL(2,\C)$ down to the compact loop gravity gauge group $\SU(2)$. This compactification, which can be understood as the origin of the discrete spectra for areas and volumes, leads to a periodicity in the (extrinsic) curvature, that one  can not fully reconstruct the metric from sampling the holonomy of the Ashtekar-Barbero connection, or in equivalent words, that the choice of observables in loop quantum gravity for a fixed Immirzi parameter does not allow to distinguish all  points of the classical phase space. In that sense, the Immirzi parameter quits being a mere coupling constant but appears to play the new effective role of a cut-off, similarly to the energy scale cut-off in usual quantum field theory. This is consistent with the view that it determines the size of discrete quanta of geometry and points towards the perpective that the bare theory would in the ``continuum limit'' $\beta \rightarrow 0$. Then specific physical situations will require specific values of the Immirzi parameter, which will determine the suitable truncation of the effective corrections to general relativity (resulting from loop quantum gravity) to use in that case.
\footnotetext{To be more precise, we can put physical dimensions back in the game and we are actually dealing with a dimensionfull Immirzi parameter, combined with the Planck unit area, $\beta\,l_{P}^{2}$.}

\medskip

We start by a very short review of the loop quantum gravity formalism, focusing of the definition of the Ashtekar-Barbero connection and its holonomy. We underline that it carries some non-vanishing torsion, proportional to the Immirzi parameter and encoding the extrinsic curvature of the canonical slice (and not an actual torsion of the 3d intrinsic geometry). The second section is devoted to the calculation of the Wilson loops of the Ashtekar-Barbero connection in a space-like 3-hyperboloid in flat space-time. We show to which extent one can recover the curvature $\ka$ of the hyperboloid from the value of a Wilson loop, which we apply to the topic of coarse-graining loop quantum gravity data. We insist on the periodicity of the Wilson loop in the curvature $\ka$ due to the Immirzi parameter. We discuss in the third section the possibility to remedy this by using a full network of Wilson loops, i.e. a full spin network state, and show that we still miss high curvature fluctuations as long as we use a locally finite graph. Finally, we conclude this short paper with the perspective of dealing with the Immirzi parameter as a cut-off for quantum general relativity.

\vfill

\section{Ashtekar-Barbero connection, Extrinsic curvature and Holonomy}

Loop quantum gravity aims at a canonical quantization of general relativity. We start with the first order Palatini action, in terms of the vierbein 1-form $e^I_{\mu}$ and the Lorentz connection $\om^{IJ}_{\mu}$, with the additional Holst term \cite{Holst:1995pc}:
\be
S[e,\om]=\int_{\cM} \star( e \wedge e)_{IJ} \wedge F[\omega]^{IJ}
-\f1\beta e_I \wedge e_J \wedge F[\omega]^{IJ}
\ee
with the curvature $F[\omega] = d\omega +\omega \wedge \omega$ or explicitly:
\begin{equation}
F[\omega]^{IJ}_{\mu \nu} = \partial_\mu \omega_\nu^{IJ} - \partial_\nu \omega_\mu^{IJ} + \delta_{KL} (\omega_\mu^{IK} \omega_\nu^{LJ} - \omega_\nu^{IK} \omega_\mu^{LJ})
\end{equation}
The new term, whose coupling constant is the Immirzi parameter $\beta$, doesn't affect the classical equations of motion as long as the tetrad $e$ is invertible and the resulting 4-metric non-degenerate.
This term can be interpreted as a mass term for the torsion, through the Nieh-Yan topological invariant \cite{Freidel:2005ak,Freidel:2005sn}.

In the canonical quantization program, one consider a globally hyperbolic spacetime $\cM\sim\R\times \Sigma$ and proceeds to the 3+1 splitting and Hamiltonian analysis of the gravity action. As reviewed in \cite{thiemann}, we get a canonical pair of field variables, the densitized triad $E$, defining the intrinsic 3d geometry of the canonical hypersurface, and the extrinsic curvature $K$, which can be understood as the ``speed'' of the 3d geometry. Assuming the time-gauge $e_0^i = \delta_0^i$, these are defined as\footnote{The variables can be defined without the time-gauge but expressions are more involved. The densitized triad diagonalizes the 3d-metric and the extrinsic curvature is defined as $K_{ab} = q_a^c q_b^d \nabla_c n_d$ in terms of the normalized time normal $n$ and where $q$ is the projection operator to the tangent space of the 3d slice defined by $q_{ab} = g_{ab} - n_a n_b$. The triad in the 3d slice is then used to lift to tangent indices.}:
\begin{equation}
\left\{\begin{array}{rcl}
E^a_i &=& \frac{1}{6} \epsilon^{abc} \epsilon_{ijk} e_b^j e_c^k \\
K_a^i &=& \omega_a^{0i}
\end{array}\right.
\end{equation}
where $a$ are space coordinates indices and $i$ are $SU(2)$ labels.

They satisfy the following constraint by definition:
\be
\forall i,~ \mathcal{G}_i = \epsilon_{ij}^{~~k} K_a^j E^a_k=0
\ee
The Poisson bracket is:
\be
\{K_a^i(x),E^b_j(y)\}
\,=\,
8\pi G \delta_a^b\delta^i_j\delta(x-y)
\ee
The Ashtekar-Barbero variables for loop gravity are introduced by a simple canonical transformation on these variables, in order to recover the phase space of a gauge field theory. We define the Ashtekar-Barbero connection from the spin-connection compatible with the triad $E$:
\be
A_a^i=\Gamma[E]_a^i +\beta K_a^i\,,
\ee
with the spin connection $\Gamma$ satisfying the covariant derivative $\nabla_a E^b_i = e (\partial_a e^b_i - \Gamma^c_{ab}e^i_c + \epsilon_{ij}^{~~k} \Gamma_a^j e_k^b)= 0$, where $e$ is the determinant of cotriad $(e^i_a)$. $\Gamma^c_{ab}$ is the torsionless (symmetric in low indices) affine connection compatible with the 3-metric $h_{ab} = e_a^i e_b^j \delta_{ij}$. The compatibility condition is expressed $\nabla_a h_{bc} = 0$. Putting everything together, the spin connection reads explicitly:
\be
\begin{array}{rcl}
\Gamma[E]_a^i &=& \frac{1}{2}\epsilon^{ijk}E^b_j\left(\partial_a E^j_b - \partial_a E^j_b + E^c_j E^\ell_a \partial_b E^\ell_c\right) \\
&+& \frac{1}{4}\epsilon^{ijk}E^b_k\left(2E^j_a \frac{\partial_b \det E}{\det E} - E^j_b \frac{\partial_a \det E}{\det E}\right).
\end{array}
\ee
The second term involving $\det E = \epsilon^{ijk} \epsilon_{abc} E_i^a E_j^b E^k_c = e^2$ comes from the fact that we are dealing with the densitized triad. The Poisson bracket still reads:
\be
\{A_a^i(x),E^b_j(y)\}
\,=\,
8\pi G \beta \delta_a^b \delta^i_j \delta(x-y)
\ee
while the orthogonality constraint between $E$ and $K$ now becomes a Gauss law:
\be
(D_{A}E)_i = \partial_a E^a_i + \epsilon_{ij}^{~~k} A_a^j E_j^a
=0\,.
\ee
This can be seen directly from the compatibility condition between the triad and the spin-connection and comes from the more general condition $\nabla_a e^b_i =0$ (written earlier using the densitized triad). We underline that this equation, and thus the Gauss law, holds whatever the value of the Immirzi parameter $\beta$. The difference between the Ashtekar-Barbero connection and the spin-connection then resides in that fact that the connection $A$ carries a non-vanishing torsion $T[e,A]$, which is proportional to the Immirzi parameter $\beta$ and reflects the extrinsic curvature:
\be
\begin{array}{rcl}
T[e,A]^i_{ab}&\equiv&
(d_A e^i)_{ab} \\
&=& \partial_a e^i_b - \partial_b e^i_b + \epsilon^i_{~jk} (A^j_a e^k_b - A^j_b e^k_a) \\
&\,=\,&
\beta \epsilon^i_{~jk} (K^j_a e^k_b - K^j_b e^k_a) \\
&\,=\,& \beta \epsilon^i_{~jk} (K^j \wedge e^k)_{ab}
\end{array}
\ee
In some sense, we can interpret the Gauss law $D_{A}e=0$ as the longitudinal part of the parallel transport of the triad $e$ by the Ashtekar-Barbero connection while $T=d_{A}e=\beta K\w e$ is its transversal part.

\medskip

After the choice of the Ashtekar-Barbero variables to describe the space-time geometry, the second prescription of loop quantum gravity is the choice of a specific set of observables, forming the holonomy-flux algebra \cite{Lewandowski:2005jk}. More precisely, we consider a class of cylindrical functionals of the connection, which generalizes the Wilson loop and depend on the holonomies of the connection $A$ along finite sets of edges within the canonical hypersurface $\Sigma$. Let us start with the Wilson loop. We consider a closed loop $\gamma$ and consider the gauge-invariant observable defined by the trace of the holonomy around that loop:
\be
W_{\gamma}[A]
\,\equiv\,
\tr\,U_{\gamma}[A]
\,=\,
\tr\,\cP e^{\int_{x \in \gamma}\,A_a^i J_i \left(\frac{\mathrm{d}x}{\mathrm{d}s}\right)^a},
\ee
where the $J_i$ are the generators of the $\mathfrak{su}(2)$ Lie algebra. In the fundamental representation, $J_i = \frac{\sigma_i}{2}$ are the Pauli matrices. Of course, one must in principle choose a starting point, or root, for the loop, but the trace does not depend on that choice in the end.
More generally, we consider a closed connected oriented finite  graph $\cG$ and construct a cylindrical functional of the connection $A$ which depends only on the holonomies of $A$ along the edges $e$ of the graph:
\be
\Psi_{\cG}[A]
\,\equiv\,
\psi\left(
\{U_{e}[A]\}_{e\in\cG}
\right)\,
\ee
such that the function $\psi$ is invariant under gauge transformations, which act by $\SU(2)$ transformations at the graph vertices:
\be
\psi\left(
\{U_{e}\}_{e\in\cG}
\right)
=
\psi\left(
\{h_{s(e)}U_{e}h_{t(e)}^{-1}\}
\right),
\,
\forall h_{v}\in\SU(2)
\ee
These cylindrical wave-functions realize a sampling of the geometry at the classical level (see e.g. \cite{Freidel:2011ue}), but they will entirely define the state of geometry at the  quantum level.
Considering holonomies of the Ashtekar-Barbero connection for a fixed real Immirzi parameter irremediably introduces a periodicity in the extrinsic curvature $K$, which erases all the information about its high momentum fluctuations. Moreover the intrinsic curvature and the extrinsic curvature are also mixed in the  Ashtekar-Barbero holonomies and it becomes a challenge to use the triad (or its discretized version as fluxes living of the graph edges) to distinguish the two contributions, intrinsic and extrinsic, to the connection in order to faithfully reconstruct the space-time geometry (at least around the canonical hypersurface).  This is especially important when investigating the renormalization flow of loop quantum gravity as resulting from the coarse-graining of its quantum geometry.

\medskip

The two important points that we would like to underline in this short paper are:
\begin{itemize}
\item The connection $A$ is not a space-time connection, except in the special case of the (anti-)self-dual Ashtekar connection ${\cal A}$ for the purely imaginary choice $\beta=\pm i$. It depends on the space-time embedding of the canonical hypersurface $\Sigma$. Considering a Wilson loop $\gamma$, its value will change if we embed it in different canonical space-like hypersurfaces $\Sigma$. This apparently bad feature might be turned into an advantage: it could help us extract a a simpler way the extrinsic data from the Ashtekar-Barbero holonomies and reconstruct the local embedding of the 3d space manifold. 

\item The  Ashtekar-Barbero connection, for real $\beta$, is in some sense a projection of the non-compact Lorentz connection into the compact $\SU(2)$ group. We lose some information, due to the periodicity in the extrinsic curvature. At the classical level, different extrinsic curvatures will still lead to the same value of the Wilson loop. This appears to impose a cut-off on the possible excitations of the geometry, more precisely on the extrinsic curvature, i.e. on the speed/momentum of the 3d intrinsic geometry. 

\end{itemize}

We will illustrate these two points in the following sections with the example of a closed loop embedded in space-like hyperboloids with variable curvature within the flat 4d space-time. We will discuss the dependence of the Wilson loop on the curvature of the hyperboloid, to show both how the Ashtekar-Barbero connection depends on the space-time embedding and how we can recover the extrinsic curvature from the value of the holonomy.

\section{Wilson loops on the Hyperboloid}
\label{WilsonLoops}

\begin{figure}[h]
\includegraphics[height=70mm]{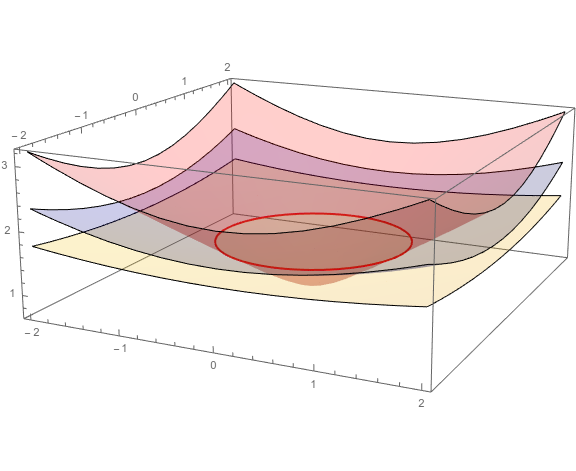}
\caption{\label{plot3d} This shows several hyperboloids of different curvature all containing the same loop in flat spacetime. The curvature of the embedding hyperboloid affects however the curvature of the Ashtekar-Barbero connection.}
\end{figure}

Let us start with  the flat 3+1d Minkowski space-time with signature (-+++) and consider the upper sheet of the space-like hyperboloid,
\be
-(t-t_{0})^{2}+(x^{2}+y^{2}+z^{2})=-\ka^{2},\quad
t\ge t_{0}\,,
\ee
with an arbitrary curvature radius $\ka>0$ and a possible time shift $t_{0}\in\R$. We would like to look at the Ashtekar-Barbero holonomy around a loop of radius $R$, say
\be
\gamma\equiv\{t=T,\,x^{2}+y^{2}+z^{2}=R^{2}\}\,,
\ee
where $T$ and $R$ are arbitrarily fixed.
As illustrated on fig.{\ref{plot3d}}, we embed this loop in the whole family of hyperboloid of arbitrary curvature radius $\ka$ by adjusting their time shift in terms of $\ka$,
\be
t_{0}=T-\sqrt{R^{2}+\ka^{2}}\,.
\ee
This setting is very similar to \cite{Samuel:2000ue}, but we extend that calculation explicitly to arbitrary curvature $\ka$.

\medskip

\begin{figure}[h]
\includegraphics[height=40mm]{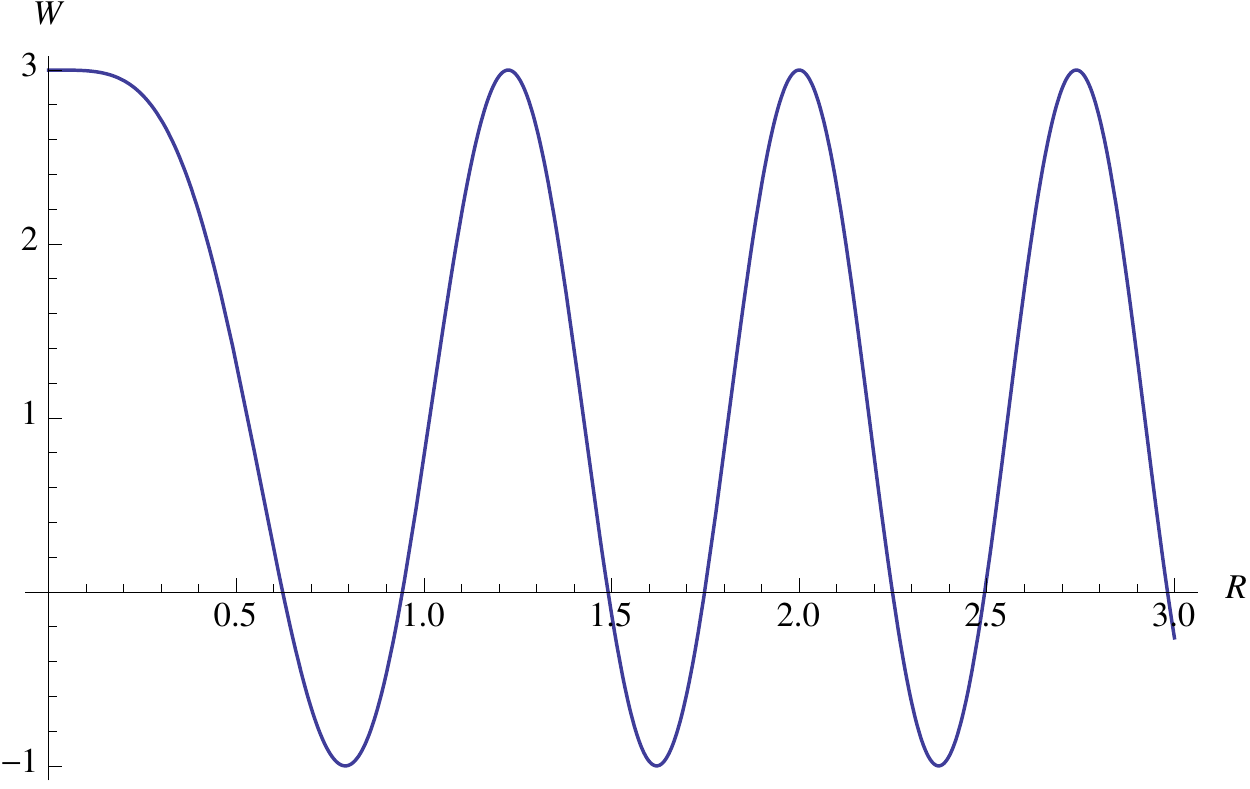}
\includegraphics[height=40mm]{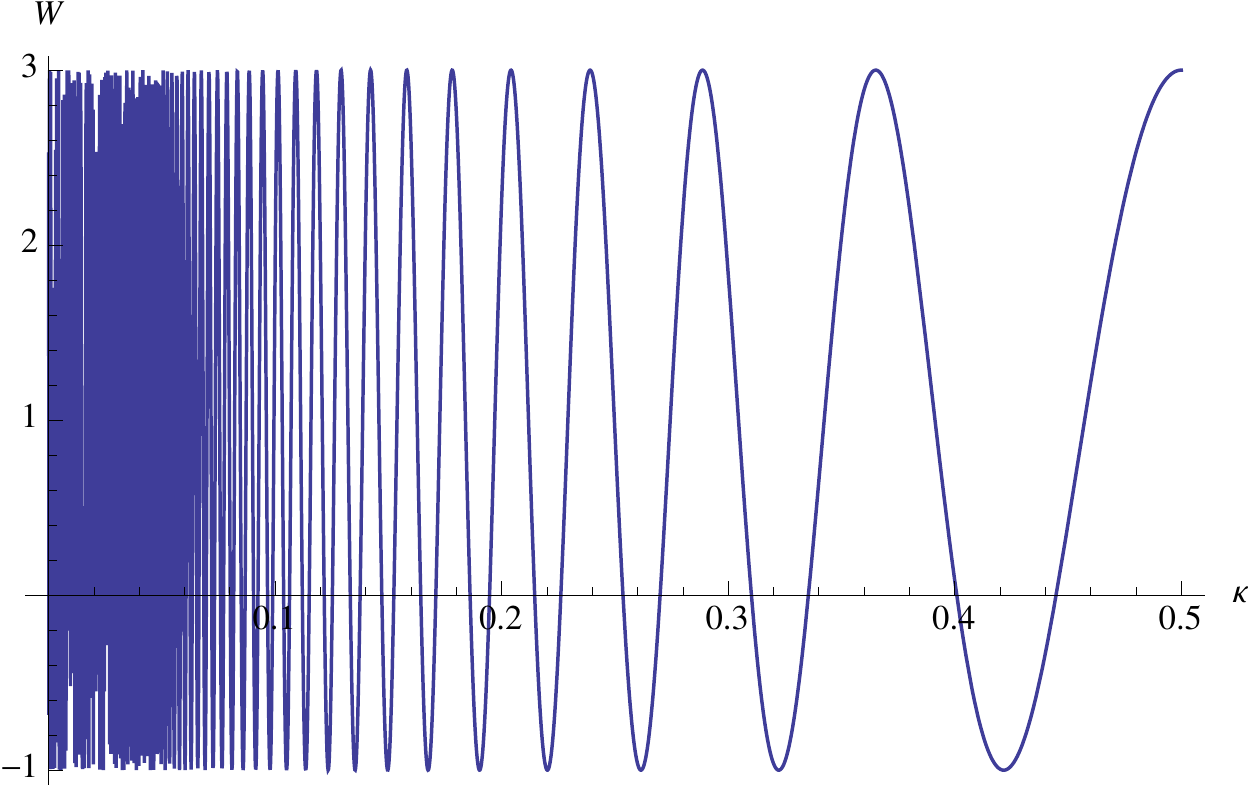}
\caption{\label{W}The Wilson loop $W$ plotted in terms of the loop size $R$ (in units of $\ka$) in the upper graph, and in terms of the curvature radius $\ka$ (in units of $R$) in the lower graph, both for a Immirzi parameter set to $\beta=1$.}
\end{figure}

Let us now compute the Ashtekar-Barbero connection of the hyperboloid and the value of the Wilson loop in terms of $R$, $\ka$ and the Immirzi parameter $\beta$.
Using the spherical coordinate on the hyperboloid, the canonical vector basis is:
\be
\begin{array}{rcl}
\partial_{r} &=& \sin\theta\cos\varphi \partial_x + \sin \theta \sin \varphi \partial_y + \cos \theta \partial_z + \frac{r}{\sqrt{\kappa^2 + r^2}}\partial_t \\
\partial_{\theta} &=& r\cos\theta\cos\varphi \partial_x + r\cos \theta \sin \varphi \partial_y - r\sin \theta \partial_z \\
\partial_{\varphi} &=& -r \sin\theta\sin\varphi \partial_x + r\sin \theta \cos \varphi \partial_y
\end{array}
\ee
the triad field is:
\be
\begin{array}{rcl}
\hat{e}_1 &=& \sqrt{1 + \frac{r^2}{\kappa^2}}\partial_r \\
\hat{e}_2 &=& \frac{1}{r}\partial_\theta \\
\hat{e}_3 &=& \frac{1}{r\sin \theta}\partial_\varphi \\
\end{array}
\ee
This allows to compute the affine connection:
\be
\Gamma^{a}_{bc} = \frac{1}{2}h^{ad}(\partial_b h_{dc} + \partial_c h_{db} - \partial_d h_{bc})
\ee
where $h$ is the metric. This gives the spin-connection:
\be
(\Gamma^{i}_{a}) = \begin{pmatrix}
0 & 0 & 0 \\
0 & 0 & - \sqrt{1+\frac{r^2}{\kappa^2}} \\
- \cos \theta & - \sin \theta \sqrt{1+\frac{r^2}{\kappa^2}} & 0
\end{pmatrix}
\ee
where $a$ is the row number and $i$ labels the column. We also compute the extrinsic curvature $K^{i}_{a}$, which describes the variation of the time-normal to the hypersurface (projected onto it):
\be
(K^{i}_{a}) = \begin{pmatrix}
\frac{1}{\kappa} \sqrt{1+\frac{r^2}{\kappa^2}} & 0 & 0 \\
0 & \frac{r}{\kappa} & 0  \\
0 & 0 & \frac{r}{\kappa}\sin \theta
\end{pmatrix}
\ee
These are combined into the Ashtekar-Barbero connection $A^{i}_{a}=\Gamma^{i}_{a}+\beta K^{i}_{a}$,
\be
(A^{i}_{a}) = \begin{pmatrix}
\beta \frac{1}{\kappa} \sqrt{1+\frac{r^2}{\kappa^2}} & 0 & 0 \\
0 & \beta \frac{r}{\kappa} & - \sqrt{1+\frac{r^2}{\kappa^2}} \\
- \cos \theta & - \sin \theta \sqrt{1+\frac{r^2}{\kappa^2}} & \beta \frac{r}{\kappa}\sin \theta
\end{pmatrix}
\ee
It is then straightforward to compute its holonomy around the loop $\gamma$, the interested reader will find the detailed calculation in the appendix \ref{holonomy}. We obtain for the spin-1 Wilson loop (for the 3-dimensional representation, where the holonomy is represented as a $\SO(3)$ group element):
\be
W_{\ka}(R)=1+2\cos\left(
2\pi\sqrt{1+(1+\beta^{2})\f{R^{2}}{\ka^{2}}}
\right)
\ee
We see a clear dependence of the size of the loop in units of the curvature radius of the hyperboloid, as illustrated on the plots in fig.\ref{W}.

This term further depends on the Immirzi parameter $\beta$.
For $\beta^{2}=-1$, this extra term vanishes and we recover $W_{\ka}(R)=3$, which signals a flat connection. This is indeed the case for the complex (anti-)self dual Ashtekar connection, which is a space-time connection and sees that the initial space-time here is flat.
However in general, the Ashtekar-Barbero connection is not flat, even though the space-time is flat, and contains information about the curvature $\ka$ of the hyperboloid. Thus, we can turn that apparently bad feature that $A$ is not a space-time connection into the advantage that we can retrieve some information about the curvature of the canonical slice from the value of the Wilson loop.
Indeed, one can invert the relation above and obtain the dimensionless ratio $\ka/R$ from $W$.

This leads us to the following proposition, useful in the context of coarse-graining loop quantum gravity:

\begin{prop}
In the context of the coarse-graining of loop quantum gravity, let us consider a Wilson loop of size $R$ small enough compared to the typical curvature radius. Then assuming by the equivalence principle that the 4d metric is locally flat in the considered space-time region, and assuming the homogeneity of the local 3d geometry, we can derive the 3d curvature radius $\ka$ from the value of the Wilson loop $W$ in units of the loop size, for a fixed real value of the Immirzi parameter $\beta\in\R$:
\be
\f{\ka^{2}}{R^{2}}
=
\f{1+\beta^{2}}{(\varphi+k)^{2}-1},
\,\,k\in\Z
\ee
with the angle $\varphi$ given in terms of the Wilson loop by:
\be
2\pi\varphi
=
\cos^{-1}\left(
\f{W-1}2
\right)
\,\,\in[0,\pi]\,.
\ee
The curvature is not uniquely fixed but determined up to a period $k\in\Z$.
\end{prop}

The periodicity implies an ambiguity in the determination of the curvature from the Wilson loop. One could  decide to take the lowest value of the curvature, i.e the highest value of the curvature radius, typically given by the natural choice $k=1$. But this would mean obviously neglecting the possibility of higher curvature fluctuations. In this sense, we see that fixing a real Immirzi parameter leads to a cut-off in curvature in the context of loop quantum gravity.


\medskip

Here we have assumed a flat space-time and considered a homogeneous space slice. For later investigation, it would be interesting to extend this analysis to a homogeneous curved space-time with non-vanishing cosmological constant $\Lambda \ne 0$.  Technically we would then have two length ratios  to determine, the space curvature and the space-time curvature in terms of the Wilson loop size. Physically, this would allow to distinguish the intrinsic and extrinsic curvature of the canonical hypersurface and help clarifying the relationship between the  torsion and the extrinsic curvature in loop quantum gravity. Finally, from the perspective of coarse-graining, it is necessary to see how the fluctuations of gravity renormalize the scalar curvature.

\medskip

Moreover, we have focused on the Wilson loop observable, measuring the curvature of the Ashtekar-Barbero connection. In light of the interplay between curvature and torsion in loop quantum gravity, it might be interesting to also look at an observable discretizing the torsion, such as a surface integral $\iint_{\cS} K\w e$ on the minimal surface supported by the loop. Having such an observable at our disposal would allow a more direct access to the extrinsic curvature. We think that this might be related to the holomorphic holonomy of the spinorial formalism for loop gravity \cite{Livine:2013zha,Livine:2011gp,Livine:2011vk,Dupuis:2012vp}. This link will be investigated in future work.

\section{Reconstructing the Geometry From Holonomies}

\begin{figure}[h]
\includegraphics[height=40mm]{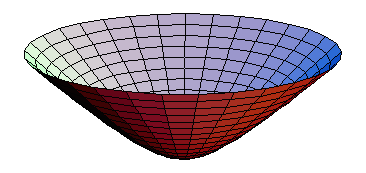}
\caption{\label{mesh} This shows a network of loops on a 2d hyperboloid, with increasing loop size and transversal links between them creating a grid. When moving up to a 3d hyperboloid, the third dimension will make it harder to locate a loop ``inside'' or ``outside'' another one.}
\end{figure}

Up to now, we have looked at a single Wilson loop. Of course, it would get much more information if we were to consider a mesh of Wilson loops and have several samples of the Ashtekar-Barbero connections through the holonomies around various loops. Having access to loops of various sizes $R_{1}, R_{2},..,R_{n}$ instead of a single loop unit $R$ would surely be very helpful. We could introduce a whole network graph, with loops of increasing sizes  as illustrated on fig.\ref{mesh},  on our canonical slice and consider all the data living on the corresponding spin network state.

We see three (inter-related) short-comings of this scenario:
\begin{itemize}
\item This very much resembles a discretization scheme and, if we don't assume that the canonical slice is homogeneous on the whole graph, we know that this will not pick up the curvature fluctuations with momentum higher than the network spacing.

\item We can not assume to know all the loop sizes   $R_{1}, R_{2},..,R_{n}$. This would already mean assuming a lot of information on the geometry of the hypersurface! Working with a single loop, we don't assume knowing its size $R$ and we merely use it as a length unit for dimensionfull observables.

\item We can not locate a priori the loops with respect to each other, first, because we do not have an assumed background geometry to do so and evaluate the distances and so on, and, second, because we are actually considering equivalence of graph under diffeomorphisms. So there is no a priori notion of which loop is inside or smaller than another one and this has to be determined a posteriori from the geometry reconstruct from the spin network state itself.

\end{itemize}

The natural way to sidestep these obstacles is to introduce extra data, such as a sampling of the triad or torsion or extrinsic curvature and not only of the Ashtekar-Barbero connection. The natural arena for this is the twisted geometry framework and we will investigate in the future how to define further gauge-invariant observables probing the extrinsic curvature and space-time embedding of the canonical hypersurface in order to describe the space-time geometry around the spatial slice.

\section{The Immirzi parameter as a cut-off for Quantum General Relativity}

\begin{figure}[h]
\includegraphics[height=40mm]{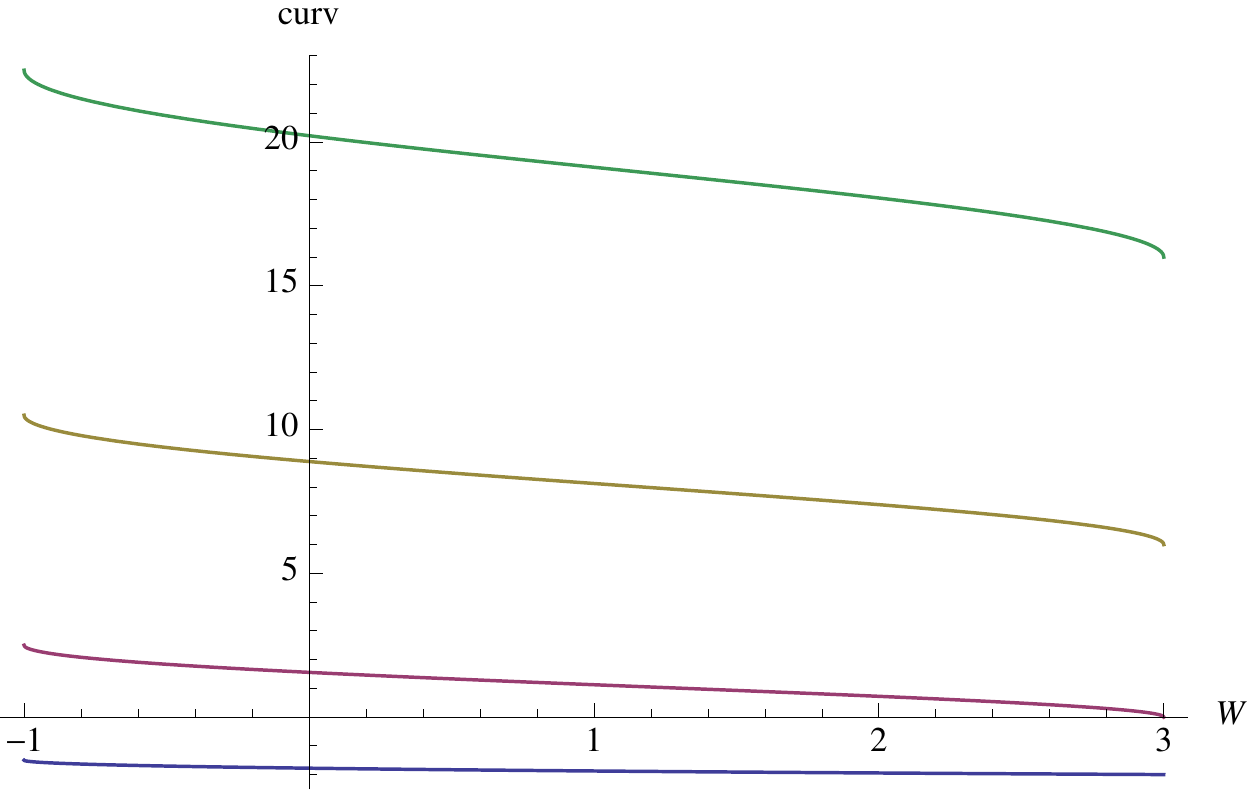}
\caption{\label{curvfromW}Curvature $R^{2}\ka^{-2}$ from the Wilson loop values for values of $k=0..3$  for a Immirzi parameter set to $\beta=1$: different values of $k$ lead to explore different sectors for the curvature.}
\end{figure}

Focusing the value of the Wilson loop and the inversion formula to recover the curvature, let us push further the consequences of the boundedness of the Wilson loop for a real Immirzi parameter $\beta\in\R$ and its associated periodicity in the curvature:
\be
R^{2}\ka^{-2}
=
\f1{1+\beta^{2}}\left[
\left(k+\f{\cos^{-1}\f{(W-1)}2}{2\pi}\right)^{2}-1
\right]\,,
\ee
with $k\in\Z$. Different values of the periodicity parameter $k$ define different slices for the curvature $\ka$, as we show with the plots on fig.\ref{curvfromW}
For the value giving the minimal curvature sector, $k=\pm 1$, we have a simple bound of the curvature $\ka^{-1}$:
\be
R^{2}\ka^{-2}
\le
\f1{1+\beta^{2}}\,\f54
\ee
Therefore, if we naturally choose a single section for this inversion formula to uniquely determine the curvature from the Wilson loop, we do have a cut-off in the possible curvature.

This suggests a slight shift in perspective on the role of the Immirzi parameter in loop quantum gravity: it is not merely a flowing coupling constant to be renormalized, but it becomes the ``energy'' cut-off itself for the quantum theory. Indeed, it controls the maximal excitations of the extrinsic curvature, i.e. of the conjugate momentum to the intrinsic geometry.

\section*{Conclusion \& Outlook}

We have discussed the role of the Immirzi parameter $\beta$ in loop quantum gravity. It  plays at least two distinct roles. At the classical level, it controls the torsion of the Ashtekar-Barbero connection, which encodes the extrinsic curvature and thus the space-time embedding of the canonical hypersurface. However, through the choice of the holonomy observables, it further leads to a compactification of the gauge group (from the non-compact Lorentz group to the compact $\SU(2)$ group), at least for a real Immirzi parameter $\beta\in\R$. This translates into a periodicity of the observables, on which the quantization is based, on the extrinsic curvature. This underlines the role of the Immirzi parameter as a cut-off for the general relativity phase space and for the quantum fluctuations of the geometry: the limit $\beta\rightarrow\infty$ suppresses curvature fluctuations, while the limit $\beta\rightarrow0$ allows to cover the whole phase space. This is very similar to the role of the energy cut-off in the renormalisation of quantum field theory. At the mathematical level, it implies looking at the renormalization flow of the other coupling constants (such Newton's gravity constant and the cosmological constant) in terms of $\beta$. And at a more physical level, it means that one should adjust the Immirzi parameter according to the physical events for which one is trying to make predictions.

We illustrated this with the simple calculation of the Wilson loop of the Ashtekar-Barbero connection on a fixed loop living in various space-like hyperboloid embedded in the flat space-time and discussed how to invert the relation between the hyperboloid curvature and the value of the Wilson loop. Our computation is also relevant in the context of the coarse-graining of loop quantum gravity, when the goal is to evaluate the average curvature carried by the geometry fluctuations and attempting to reconstruct the mean geometry at large scale.

A direct possible technical improvement of our result would be to take into account the space-time curvature, as a non-vanishing cosmological constant, and not only the space curvature. This would allow to distinguish the intrinsic and extrinsic curvature of the spatial slice in the reconstruction process of the geometry from the algebraic data carried by the quantum states of geometry in loop quantum gravity. Another direction of investigation will be to look for discretized observables encoding in a more direct fashion the Ashtekar-Barbero torsion and one potential path we would like to explore in the future is the holomorphic holonomy defined in the spinorial formalism for loop gravity.

An interesting possibility to check the role of the Immirzi parameter as a cut-off in a renormalization scheme is to test it in usual Quantum Field Theory. Indeed, loop-like quantization schemes for scalar fields have been developed \cite{Ashtekar:2002vh,Laddha:2010hp}. In this framework, the quantization is defined with an Immirzi-like parameter constraining the momentum of the scalar field to live in $\U(1)$ rather than $\R$ and thus acting as a cut-off in the field modes. The flow of renormalization with respect to this cut-off could be studied and compared with the standard renormalization flow in terms of an energy cut-off.

Finally, we would like to conclude this short letter with a speculation on the consequences of the Immirzi parameter in loop quantum gravity: similarly to the cosmological constant $\Lambda>0$, which plays the role of an infrared cut-off and is taken into account through a $q$-deformation of the gauge group (both of the Lorenz group and of the $\SU(2)$ group) (see e.g. \cite{Noui:2002ag,Haggard:2014xoa,Rovelli:2015fwa,Dupuis:2013haa,Dupuis:2013lka}), we propose that the Immirzi parameter $\beta\in\R$, which  might be related to a quantum deformation (or central extension) of the space-time diffeomorphism algebra (or its discrete quantum equivalent acting on loop quantum gravity's spin network states).

%

\appendix

\section{Calculating the Wilson loop on the hyperboloid}
\label{holonomy}

We want to compute the Ashtekar-Barbero holonomy in space-time flat around a loop defined by:
\begin{equation}
\left\{\begin{array}{rcl}
t &=&  T \\
x &=& R \cos \phi \\
y &=& R \sin \phi \\
z &=& 0
\end{array}\right. , \qquad \phi \in [0,2\pi[
\end{equation}
$\phi$ is the coordinate along the circle. Note here that $t$ is constant set to $T$ and is irrelevant in the following calculation. We embed this loop on the hyperboloid defined by $(t-t_0)^2 - x^2 - y^2 - z^2 = \kappa^2$ using $t_0 = T - \sqrt{\kappa^2 + R^2}$. We use the spherical coordinates on the hyperboloid and compute all the fields and the Ashtekar-Barbero connection as explained in the main text in section \ref{WilsonLoops}. To compute the holonomy around the loop, we contract the connection with the tangent vector along the circle given by:
\begin{equation}
(t^a) = \frac{\mathrm{d} x^a}{\mathrm{d} \phi} = (0 ~ 0 ~ 1)^T
\end{equation}
We can then compute the connection along the circle $A = t^a A^i_a \frac{\sigma_i}{2}$:
\begin{equation}
A = \begin{pmatrix}
0 & \frac{1}{2}\left(\sqrt{\frac{\kappa^2 + R^2}{\kappa^2}} -i\beta \frac{R}{\kappa} \right) \\
\frac{1}{2}\left(\sqrt{\frac{\kappa^2 + R^2}{\kappa^2}} +i\beta \frac{R}{\kappa} \right) & 0
\end{pmatrix}
\end{equation}
where the $\sigma_i$ are the Pauli matrices. The holonomy $H(\phi)$ along the arc of the circle $(0,\phi)$ is then computed through:
\begin{equation}
\frac{\mathrm{d}H}{\mathrm{d} \phi} = i A H(\phi)~,\quad H(0) = \id
\end{equation}
This differential equation is integrated into:
\begin{equation}
H(\phi) = \cos (\alpha \phi) + \sin (\alpha \phi) \frac{A}{\alpha}
\end{equation}
where $\alpha = \sqrt{- \det A} = \frac{1}{2}\sqrt{\frac{\kappa^2 + R^2}{\kappa^2} + \beta^2 \frac{R^2}{\kappa^2}}$.

The measurable quantity (that is the gauge invariant quantity) is the trace of the closed holonomy which is $2\cos 2\pi\alpha$. However, our choice of coordinates is degenerate along the line $\theta \equiv 0 (\pi)$. As a consequence, the loop is not contractible with this choice of coordinates and after one loop, in the flat case, we don't get the identity but its opposite. We avoid this problem by using the vector representation of $SU(2)$ and take the trace in this representation which does not see this sign. We get:
\begin{equation}
W_\kappa(R) = 4\cos^2\left(2\pi\alpha\right) - 1
\end{equation}
which we finally write:
\begin{equation}
\boxed{W_\kappa(R) = 1+ 2\cos\left(2\pi\sqrt{1 + (1 + \beta^2)\left(\frac{R}{\kappa}\right)^2}\right).}
\end{equation}

~


\bibliographystyle{bib-style}
\bibliography{lqg}

\end{document}